\documentclass[conference]{IEEEtran}
\IEEEoverridecommandlockouts
\usepackage{cite}
\usepackage{amsmath,amssymb,amsfonts}
\usepackage{algorithmic}
\usepackage{graphicx}
\usepackage{textcomp}
\usepackage{xcolor}
\usepackage{algorithm}
\usepackage{algorithmic}
\def\BibTeX{{\rm B\kern-.05em{\sc i\kern-.025em b}\kern-.08em
    T\kern-.1667em\lower.7ex\hbox{E}\kern-.125emX}}
\begin{document}

\title{Robust Particle Filtering via Bayesian Nonparametric Outlier Modeling
\thanks{This work was partly supported by National Natural Science Foundation
(NSF) of China (No. 61571238), Scientific Research Foundation of Nanjing
University of Posts and Telecommunications (No.NY218072) and a research
fund from Yancheng Big Data Research Institute.}
}

\author{\IEEEauthorblockN{Bin Liu}
\IEEEauthorblockA{\textit{School of Computer Science} \\
\textit{Jiangsu Key Lab of Big Data Security $\&$ Intelligent Processing} \\
\textit{Nanjing University of Posts and Telecommunications}\\
Nanjing, China \\
bins@ieee.org}
}

\maketitle

\begin{abstract}
This paper is concerned with the online estimation of a nonlinear dynamic system from a series of noisy measurements. The focus is on cases wherein outliers are present in-between normal noises. We assume that the outliers follow an unknown generating mechanism which deviates from that of normal noises, and then model the outliers using a Bayesian nonparametric model called Dirichlet process mixture (DPM). A sequential particle-based algorithm is derived for posterior inference for the outlier model as well as the state of the system to be estimated. The resulting algorithm is termed DPM based robust PF (DPM-RPF). The nonparametric feature makes this algorithm allow the data to ``speak for itself" to determine the complexity and structure of the outlier model. Simulation results show that it performs remarkably better than two state-of-the-art methods especially when outliers appear frequently along time.
\end{abstract}


\section{Introduction}\label{sec:intro}
This paper deals with the online estimation of states in nonlinear and non-Gaussian dynamic systems based on noisy measurements polluted by outliers. Particle filters (PFs), also known as Sequential Monte Carlo (SMC) methods, are mainly used for state estimation in nonlinear and non-Gaussian systems \cite{arulampalam2002tutorial,carpenter1999improved,gordon1993novel}. However, most existent PF methods in the literature adopt a pre-determined parametric model, e.g., a zero-mean Gaussian, to characterize the statistical property of the measurement noise. This simple treatment will lead to a significant degradation in filtering performance when the actual measurements are with the presence of outliers. To lessen such model mismatch problem caused by the presence of outliers, the common practice is to resort to the multiple model strategy (MMS), namely by employing multiple pre-set models together to characterize the statistical property of normal noises together with outliers \cite{liu2017robust,yi2016robust,liu2011instantaneous,Drovandi2014,urteaga2016sequential}. An efficient approach to handle model uncertainty incurred by employing multiple models is Bayesian model averaging \cite{liu2017robust}.

A limitation of the aforementioned MMS based methods is that, to use them, one has to specify a set of candidate models beforehand even if there is no prior knowledge available for model specification. To this end, an incremental learning assisted particle filtering (ILAPF) algorithm is proposed \cite{liu2018ilapf}. The basic idea underlying ILAPF is to learn an outlier model online instead of specifying a set of candidate models offline.
The ILAPF algorithm is shown to be simple while efficient, while its drawback is that it only uses a uniform distribution to roughly characterize the statistical pattern of the outliers. The uniform distribution is certainly unsatisfactory when the true distribution pattern of the outliers is much more complex and far away from being uniform. This observation motivates us to develop a more powerful learning assisted PF algorithm, which is able to reveal and then make use of any possible complex patterns in the outliers' distribution. We propose using Bayesian nonparametric DPM to model the generative mechanism of the outliers. We show that our algorithm allows the data ``speak for itself" to determine the complexity and structure of the outlier model, thus sidestepping the issue of pre-specifying candidate models and model selection.

The DPM model was recently introduced to deal with switching linear dynamical models in e.g., \cite{caron2008bayesian,magnant2015dirichlet,fox2011bayesian}, which assume that the state transition prior is uncertain. In contrast with such previous work, this work assumes that the state transition prior is precisely known, and focus on taking advantage of DPM in modeling the measurement noise.

The remainder of the paper is organized as follows. Section \ref{sec:model} succinctly describes our model. Section \ref{sec:dpm-rpf} presents the proposed algorithm in detail. Section \ref{sec:simulation} reports the simulation results, and finally, Section \ref{sec:conclusion} concludes.
\section{Model}\label{sec:model}
We consider a state space model as follows
\begin{eqnarray}\label{eqn:state-space}
\textbf{x}_t&=&f(\textbf{x}_{t-1})+\textbf{u}_t\\
\textbf{y}_t&=&h(\textbf{x}_t)+\textbf{n}_t,
\end{eqnarray}
where $t$ denotes the discrete time index, $\textbf{x}\in\mathbb{R}^{d_x}$ the state of interest to be estimated,
$\textbf{y}\in\mathbb{R}^{d_y}$ the measurement observed, $f$ the state transition function, $h$ the measurement function,
$\textbf{u}$ and $\textbf{n}$ are independent identically distributed (i.i.d.) process noise and measurement noise, respectively.
The probability density function (pdf) of $\textbf{u}_t$ is precisely known. $\textbf{n}_t$ may be a standard measurement noise or an outlier. For the former case, we have $\textbf{n}_t\sim\mathcal{N}(\mu_{(0)},\Sigma_{(0)})$, and for the latter $\textbf{n}_t\sim F(\cdot)$, where $F(\cdot)$ denotes an unknown outlier distribution. The symbol $\sim$ means \emph{distributed according to}, and $\mathcal{N}(\mu,\Sigma)$ denotes Gaussian with mean $\mu$ and covariance $\Sigma$.
Considering an outlier set $O$, wherein its elements $\textbf{o}_{(1)}, ...,\textbf{o}_{(I)}$ are statistically exchangeable, we express $F(\cdot)$ as a DPM model as follows
\begin{eqnarray}\label{eqn:dpm}
\mathbb{G}&\sim&DP(\mathbb{H},\alpha), \\\nonumber
\theta_{(i)}|\mathbb{G}&\sim&\mathbb{G}, \\\nonumber
\textbf{o}_{(i)}|\theta_{(i)}&\sim&g(\cdot|\theta_{(i)}),\nonumber
\end{eqnarray}
where $DP(\mathbb{H},\alpha)$ is a Dirichlet process (DP) parameterized by a concentration
paramter $\alpha>0$ and a base distribution $\mathbb{H}$ \cite{ferguson1973bayesian,teh2011dirichlet}, $\mathbb{G}$ is a random distribution drawn from the DP, $\theta_{(i)}\in\Theta$ is the parameter of the cluster to which $\textbf{o}_{(i)}$ belongs. Here and in what follows, the notation $(i)$ in a subscript represents the index of a data item in a set, where the bracket is used to discriminate it from the time index.
By integrating over $\mathbb{G}$, we obtain a marginal representation of the prior distribution of $\theta_{(i+1)}$ as follows
\begin{equation}
\theta_{(i+1)}|\theta_{(1)},\ldots,\theta_{(i)}\sim\frac{1}{\alpha+i}\left(\alpha\mathbb{H}+\sum_{j=1}^i\delta_{\theta_{(j)}}\right),
\end{equation}
where $\delta_{\theta}$ denotes the delta-mass function located at $\theta$. This representation is often known as the Blackwell MacQueen urn scheme \cite{blackwell1973ferguson}.
The DP can also be represented by a Chinese Restaurant Process (CRP), which describes a partition of $\theta_{(i)}$s when $\mathbb{G}$ is marginalized out \cite{orbanz2011bayesian,aldous1985exchangeability}. According to CRP, the first outlier is assigned to the first cluster, and the $i$th outlier is assigned to the $k$th cluster with probability
\begin{eqnarray}
p(z_{(i)}=k)&=&\frac{n_k}{I-1+\alpha},\;\mbox{for}\;k\leq K \\\nonumber
p(z_{(i)}=k)&=&\frac{\alpha}{I-1+\alpha},\;\mbox{for}\;k=K+1 \\\nonumber
\end{eqnarray}
where $z$ is a membership indicator, namely $z_{(i)}=k$ means $\textbf{o}_{(i)}$ belongs to cluster $k$, $n_k$ is the number of outliers included in cluster $k$. Each cluster, say cluster $k$, is defined by a parametric pdf $g(\cdot|\theta_{(k)}^{\star})$ and a prior on $\theta_{(k)}^{\star}$. Set $g(\cdot|\theta_{(k)}^{\star})\triangleq\mathcal{N}(\cdot|\mu_{(k)},\Sigma_{(k)})$, $\theta_{(k)}^{\star}\triangleq(\mu_{(k)},\Sigma_{(k)})$, and employ a conjugate Normal-Inverse-Wishart (NIW) prior for $\theta_{(k)}^{\star}$
\begin{eqnarray}\label{eqn:niw}
\Sigma_{(k)}|\kappa,W&\sim&\mathcal{IW}(\cdot|\kappa,W^{-1}),\\
\mu_{(k)}|\Sigma_{(k)},\mu_0^{\star},\rho &\sim& \mathcal{N}(\cdot|\mu_0,\Sigma_{(k)}/\rho),\nonumber
\end{eqnarray}
where $\mathcal{IW}(\cdot|\kappa,W^{-1})$ denotes an inverse-Wishart (IW) pdf parameterized by the degree of freedom $\kappa$ and the scale matrix $W$, $\mu_0^{\star}$ and $\rho$ are the other hyper-parameters of this NIW prior.
Due to conjugacy of the NIW and Gaussian, the posterior of $\theta_{(k)}^{\star}$ based on $O$ and $Z=[z_{(1)},\ldots,z_{(I)}]$ is also NIW distributed as follows \cite{murphy2007conjugate},
\begin{eqnarray}\label{eqn:niw_posterior}
p(\theta_{(k)}^{\star})&\propto& NIW(\mu_{0}^{\star},\rho,\kappa,W)\prod_{i:z_i=k}g(\textbf{o}_{(i)}|\theta_{(k)}^{\star})\\
&=& NIW(\mu_{(k)},\rho_{(k)},\kappa_{(k)},W_{(k)}),\nonumber
\end{eqnarray}
where
\begin{eqnarray}\label{eqn:niw_parameter_updata}
\mu_{(k)}&=&\frac{\rho}{\rho+n_k}\mu_{0}^{\star}+\frac{n_k}{\rho+n_k}\bar{\textbf{o}}_{(k)}\\\nonumber
\rho_{(k)}&=&\rho+n_k\\\nonumber
\kappa_{(k)}&=&\kappa+n_k\\\nonumber
W_{(k)}&=&W+R_{(k)}+\frac{\rho n_k}{\rho+n_k}(\bar{\textbf{o}}_{(k)}-\mu_{0}^{\star})(\bar{\textbf{o}}_{(k)}-\mu_{0}^{\star})^T
\end{eqnarray}
where $R_{(k)}=\sum_{i:z_i=k}(\textbf{o}_{(i)}-\bar{\textbf{o}}_{(k)})(\textbf{o}_{(i)}-\bar{\textbf{o}}_{(k)})^T$, $\bar{\textbf{o}}_{(k)}=1/n_k\sum_{i:z_i=k}\textbf{o}_{(i)}$.

In the above model, $\mu_{(0)}$, $\Sigma_{(0)}$ are deterministic and known; $\alpha$, $\kappa$, $W$, $\mu_0^{\star}$ and $\rho$ are hyper-parameters preset by the user. The other parameters will be inferred online by the algorithm described in the next Section.
\section{Algorithm}\label{sec:dpm-rpf}
In this section, we present our algorithm, DPM-RPF, for sequential inference of the state of interest $\textbf{x}_t$ based on the model presented in the above Section.
The task here is to provide a recursive solution to compute $p(\textbf{x}_t|y_{1:t})$ (or in short $p_{t|t}$), which denotes the posterior of $\textbf{x}_t$ given measurements observed up to time $t$. Note that $p_{t|t}$ can be indeed computed from $p_{t-1|t-1}$ recursively as follows \cite{arulampalam2002tutorial}
\begin{equation}\label{eqn:filter}
p_{t|t}=\frac{p(\textbf{y}_t|\textbf{x}_t)\int p(\textbf{x}_t|\textbf{x}_{t-1})p_{t-1|t-1}d\textbf{x}_{t-1}}{p(\textbf{y}_t|\textbf{y}_{1:t-1})}.
\end{equation}

The DPM-RPF algorithm starts by initializing hyper-parameters for the DPM model, specifying the particle size $J$ of the PF, drawing a set of equally weighted random samples (also called particles) $\{x_0^j,\omega_0^j\}_{j=1}^J$ from the prior $p_{0|0}\triangleq p(\textbf{x}_0)$ and initializing the outlier set $O$ to be empty. A pseudo-code to implement DPM-RPF is shown in Algorithm \ref{alg:DPM-RPF}.

Suppose that computations of DPM-RPF at time $t-1$ have been completed. We now have at hand a set of weighted samples $\{x_{t-1}^j,\omega_{t-1}^j\}_{j=1}^J$, that satisfies
\begin{equation}
p_{t-1|t-1}\simeq\sum_{j=1}^J\omega_{t-1}^j\delta_{x_{t-1}^j},
\end{equation}
and a DPM based outlier model that has $K$ active mixing components. We show in what follows how to leverage the recursion in Eqn.(\ref{eqn:filter}) to update the particle set to obtain a Monte Carlo approximation to $p_{t|t}$. The posterior of the DPM model will also be updated if a new outlier is found.
\subsection{Importance Sampling under Model Uncertainty}
To begin with, following the importance sampling principle, we draw particles $\hat{x}_t^j$, $j=1,\ldots,J$, from a proposal distribution $q(\textbf{x}_t|\textbf{x}_{t-1},\textbf{y}_{1:t})$ and then calculate the unnormalized importance weight
\begin{equation}\label{eqn:unnormalized_weight}
\hat{\omega}_t^j=\omega_{t-1}^jp(\hat{x}_t^j|x_{t-1}^j)p(y_t|\hat{x}_t^j)/q(\hat{x}_t^j|x_{t-1}^j,y_{1:t}).
\end{equation}
Set $q(\textbf{x}_t|\textbf{x}_{t-1},\textbf{y}_{1:t})=p(\textbf{x}_t|\textbf{x}_{t-1})$ as in the Bootstrap filter \cite{gordon1993novel}, then it leads to
\begin{equation}\label{eqn:weight_update}
\hat{\omega}_t^j=\omega_{t-1}^jp(y_t|\hat{x}_t^j).
\end{equation}
From Eqn.(2), we see that the likelihood in Eqn.(\ref{eqn:weight_update}), namely $p(y_t|\hat{x}_t^j)$, is defined by the pdf of $\textbf{n}_t$. We consider $K+2$ candidate pdfs of $\textbf{n}_t$, namely $\mathcal{N}(\cdot|\mu_{(k)},\Sigma_{(k)})$, $k=0,\ldots,K+1$, each corresponding to a hypothesis on the likelihood function that should be used in Eqn.(\ref{eqn:weight_update}). Let $l$ denote the hypothesis indicator, and set
\begin{equation}\label{eq:likelihood_of_l}
p_l(y_t|\hat{x}_t^j)=\mathcal{N}(y_t-h(\hat{x}_t^j)|\mu_{(l)},\Sigma_{(l)}), l=0,\ldots,K+1.
\end{equation}
As is shown, $l=0$ indicates the standard measurement noise hypothesis. If $1\leq l\leq K$, it represents a hypothesis that $\textbf{n}_t$ is drawn from one of the active mixing components of the DPM outlier model. $l=K+1$ means that $\textbf{n}_t$ is drawn from a new mixing component of DPM that may become active later. The parameter value of the new mixing component is drawn from the NIW prior presented in Eqn.(\ref{eqn:niw}). For each hypothesis $l$, its marginal likelihood is
\begin{equation}\label{eq:marglik_of_l}
L(l)\triangleq p(\textbf{y}_t|l,\textbf{y}_{1:t-1})=\sum_{j=1}^J\hat{\omega}_{t,l}^j,
\end{equation}
where $\hat{\omega}_{t,l}^j=\omega_{t-1}^jp_l(y_t|\hat{x}_t^j)$ (Note that here $\omega_{t-1}^j$ is an output at time $t-1$ of the algorithm. It is not dependant on $l$. See the next paragraph on how $\omega_{t}^j$ is calculated). The prior of the hypothesis $l$, denoted by $p_0(l)$, is proportional to the number of data points allocated into cluster $l$. Then, using Bayes theorem, we obtain the posterior probability of hypothesis $l$ as follows
\begin{equation}\label{eq:posterior_of_l}
\pi(l)=\frac{ p_0(l)L(l)}{\sum_{k=0}^{K+1}p_0(k)L(k)}, l=0,\ldots,K+1.
\end{equation}
\subsection{Model Selection and Resampling}
Now we sample a hypothesis $m$ from the posterior by setting $m=l$ with probability $\pi(l)$, $l=0,\ldots,K+1$.
Based on hypothesis $m$, we normalize the importance weights as follows
\begin{equation}\label{eq:norm_weights}
\omega_t^j=\frac{\hat{\omega}_t^j}{\sum_{a=1}^J\hat{\omega}_t^a}, j=1,\ldots,J,
\end{equation}
where $\hat{\omega}_t^j=\omega_{t-1}^jp_m(y_t|\hat{x}_t^j)$.
An optimal estimate of $\textbf{n}_t$ in terms of minimum mean squared error (MMSE) is
\begin{equation}\label{eq:mmse_noise}
\hat{n}_t=y_t-h\left(\sum_{j=1}^J\omega_t^jx_t^j\right).
\end{equation}
We allocate $\hat{n}_t$ into cluster $m$ and increments the size of cluster $m$ by 1. If $m>0$, we add $\hat{n}_t$ into $O$ and then update $Z$ correspondingly. If $m=K+1$, we activate the new mixing component with its parameter value drawn from the NIW prior in Eqn.(\ref{eqn:niw}) and then increments $K$ by 1.
To prevent particle degeneracy, a resampling procedure is adopted, which discards particles with low weights and duplicate those with high
weights. In our experiment, we selected the residual resampling method \cite{douc2005comparison,Li2015Resampling,Hol2006on}.
\subsection{Model Refinement}\label{sec:refine}
The final building block of the DPM-RPF algorithm is termed model refinement. Only if a new mixing component of
the DPM model becomes active and meanwhile the size of the updated $O$ becomes a multiple of $A$ at the current time step, we do the model refinement operation.
\begin{algorithm}[tb]
\caption{A pseudo-code to implement DPM-RPF}
\label{alg:DPM-RPF}
\begin{algorithmic}[1]
\STATE Initialization: Configure hyper-parameters $\alpha$, $\kappa$, $W$, $\mu_0^{\star}$ and $\rho$ for the DPM model; Set $K=0$; Specify the particle size $J$ of PF; Draw $x_0^j\sim p(\textbf{x}_0)$ and set $\omega_0^j=1/J$, $\forall j\in\{1,\ldots,J\}$; Initialize $O$ and $Z$ to be empty. Initialize $A$ and $B$ for the model refinement procedure.
\FOR{$t$=1,2,\ldots}
\STATE Draw $\hat{x}_t^j\sim p(\textbf{x}_t|\textbf{x}_{t-1})$, $\forall j$;
\STATE Calculate $p_l(y_t|\hat{x}_t^j)$ by Eqn.(\ref{eq:likelihood_of_l}), $\forall l\in\{0,\ldots,K+1\}$;
\STATE Calculate $L(l)$ by Eqn.(\ref{eq:marglik_of_l}), $\forall l\in\{0,\ldots,K+1\}$;
\STATE Calculate $\pi(l)$ by Eqn.(\ref{eq:posterior_of_l}), $\forall l\in\{0,\ldots,K+1\}$;
\STATE Sample $m\sim\sum_{l=0}^{K+1}\pi(l)\delta_l$, i.e., set $m=l$ with probability $\pi(l)$, $\forall l\in\{0,\ldots,K+1\}$;
\STATE Calculate $\omega_t^j, \forall j$, by Eqn.(\ref{eq:norm_weights});
\STATE Calculate the MMSE estimate of $\textbf{x}_t$: $\bar{x}_t=\sum_{j=1}^J\omega_t^j\hat{x}_t^j$.
\STATE Calculate $\hat{n}_t$ by Eqn.(\ref{eq:mmse_noise});
\STATE Allocate $\hat{n}_t$ into cluster $m$ and increments the size of cluster $m$ by 1;
\STATE If $m>0$, add $\hat{n}_t$ into $O$ and then update $Z$ correspondingly. If $m=K+1$, activate the new mixing component with its parameter value drawn from the NIW prior, see Eqn.(\ref{eqn:niw}), and then increments $K$ by 1;
\STATE Given $\{\hat{x}_t^j,\omega_t^j\}_{j=1}^J$, perform the resampling procedure, obtaining an updated particle set $\{\hat{x}_t^j,\omega_t^j\}_{j=1}^J$, in which $\omega_t^j=1/J, \forall j$;
\STATE Check the size of $O$. If it is a multiple of $A$, do the model refinement procedure as presented in subsection \ref{sec:refine}.
\STATE Output: $\bar{x}_t$.
\ENDFOR
\end{algorithmic}
\end{algorithm}

The model refinement procedure consists of running $B$ iterations of Gibbs sampling to sample from the posterior of the model parameter based on $O$ and $Z$ as follows \cite{neal2000markov}:
\begin{itemize}
\item Sample $z_{(i)}$ from
\begin{equation}
p(z_{(i)}|Z_{-i},\pi,\theta^{\star},O)\propto\sum_{k=1}^K\left[\pi_kp(o_{(i)}|\theta_{(k)}^{\star})\textbf{\mbox{I}}_{z_{(i)},k}\right],
\end{equation}
where $Z_{-i}=[z_{(1)},\ldots,z_{(i-1)},z_{(i+1)},\ldots,z_{(I)}]$, $\textbf{\mbox{I}}_{a,b}$ takes value at 1 if $a=b$, and 0 otherwise.
\item Sample $\pi$ from
\begin{equation}
p(\pi|Z,\theta^{\star},O)\propto\mbox{Dirichlet}(n_1+\alpha/K,\ldots,n_K+\alpha/K).
\end{equation}
    \item Sample each $\theta_{(k)}^{\star}$ from the NIW posterior based on $Z$ and $O$, see Eqn.(\ref{eqn:niw_posterior})-(\ref{eqn:niw_parameter_updata}).
\end{itemize}
$A$ and $B$ are constants preset by the user. The sample yielded at the last iteration is taken as the outputted parameter configuration that will be used in the next time step.
\section{Performance Evaluation}\label{sec:simulation}
We used simulated experiments to evaluate the performance of the presented algorithm. We also considered the heterogeneous mixture model based robust PF (HMM-RPF) [4] and the ILAPF [9] as competitors for performance comparison.
\subsection{Experimental setting}\label{sec:setting}
We consider a modified version of the time-series experiment presented in \cite{van2000the}. The state transition function is
\begin{equation}
\textbf{x}_{t+1}=1+\sin\left(\frac{4\pi\mbox{mod}(t+1,60)}{100}\right)+0.5\textbf{x}_t+\textbf{u}_t, 1\leq t<600,
\end{equation}
where $\textbf{x}_1$ is fixed at 1, $\textbf{u}_t\sim Gamma(3,2)$, mod($a,b$) returns the remainder after the division of $a$ by $b$. The measurement function is specified as follows
\begin{equation}\label{measure_func_simu}
\textbf{y}_t=\left\{\begin{array}{ll}
0.2\textbf{x}_t^2+\textbf{n}_t,\quad\quad\; \mbox{if}\;\mbox{mod}(t,60)\leq30 \\
0.2\textbf{x}_t-2+\textbf{n}_t,\quad \mbox{otherwise} \end{array} \right.
\end{equation}
In the simulation, to generate a measurement at $t$, a realization of $\textbf{n}_t$ is drawn with probability $P_o$ from $F=0.5\mathcal{N}(\cdot|20,0.1)+0.5\mathcal{N}(\cdot|22,0.1)$, and with probability $1-P_o$ from $\mathcal{N}(\cdot|0,0.01)$. $F$ represents the generative distribution of the outliers and the latter is the standard measurement noise distribution \emph{a priori} known. The arrival time of and the generative distribution of the outliers are invisible to the algorithms to be tested.

In the experiments, the hyper-parameters of DPM-RPF are initialized as follows: $\alpha=1$, $\mu_0^{\star}=21$, $\kappa=10$, $W=5$, $\rho=1$, $A=10$, $B=20$. The ILAPF algorithm is initialized with $\hat{lb}=10$, $\hat{ub}=90$, which represents
the initial guess for the outliers' value range. The free parameter $I$ in ILAPF is set at 20, the same as in \cite{liu2018ilapf}. The HMM-RPF algorithm is initialized in the same way as in \cite{liu2017robust}. The particle size $J$ is fixed at 200 for every algorithm involved.
\subsection{Experimental Results}
\begin{figure}[t]
\centering
\includegraphics[width=3.5in,height=2.3in]{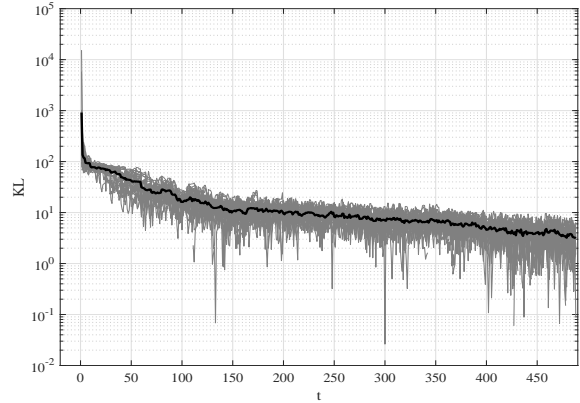}
\caption{The recorded KL distances at each time step between the posterior estimate of the outlier distribution and the real answer based on 30 independent Monte Carlo runs of the DPM based sequential outlier model inference procedure. The thick solid line represents the mean of the KL distances over those 30 runs. Note that a base 10 logarithmic scale is used for the y-axis.}\label{fig:KL}
\end{figure}

At first, we assessed the ability of DPM-RPF in discovering clustering patterns hidden in the outliers. We simulate 480 outliers drawn from $F$ and run the DPM based sequential outlier model inference part of the DPM-RPF algorithm 30 times. Fig.\ref{fig:KL} shows the Kullback-Leibler (KL) distance from the estimated and the true $F$ along time for each Monte Carlo run. It is shown that a sharp decrease in the KL distance happens at a very early stage, then the KL distance decreases gradually as more outliers appear along the time. This demonstrates that the posterior estimate of $F$ can approach the real $F$ as more and more outliers are put into the inference procedure.

\begin{figure}[t]
\centering
\includegraphics[width=3.7in,height=2.3in]{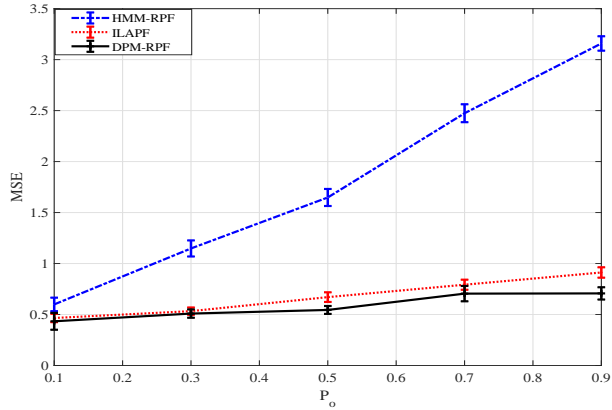}
\caption{The mean and variance of the state estimation MSE calculated over 100 independent runs of each algorithm for cases $P_o=0.1, 0.3, 0.5, 0.7$ and 0.9}\label{fig:rmse}
\end{figure}

Then we compared DPM-RPF with HMM-RPF and ILAPF in terms of the mean-square-error (MSE) of the state estimates. We calculated the mean and variance of the MSE over 100 independent runs of each algorithm. The result is plotted in Fig.\ref{fig:rmse}. As is shown, when the outliers rarely appear (corresponding to case $P_o=0.1$), DPM-RPF performs comparably with ILAPF and slightly better than HMM-RPF. As the outliers appear more and more frequently, the superiority of DPM-RPF in terms of MSE compared with its competitors becomes more and more remarkable.

\begin{figure}[t]
\centering
\includegraphics[width=3.7in,height=2.3in]{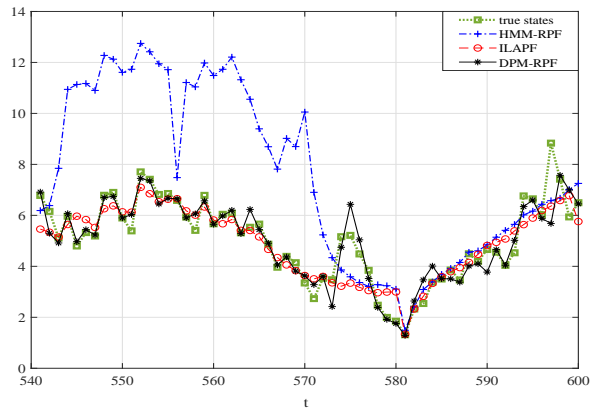}
\caption{A snapshot of the filtering result for the last 60 time steps for a frequent outlier case in which $P_o=0.9$}\label{fig:track}
\end{figure}

Fig.\ref{fig:track} shows a snapshot of the estimated trajectory of the system state yielded from an example run of the algorithms for a frequent outlier case corresponding to case $P_o=0.9$. We can see that, although the outliers appear intensively over time in the measurements (indicated by a large value of $P_o$), the DPM-RPF algorithm still works well in accurately tracking the fluctuations in the state trajectory, while ILAPF can only follow the true trajectory roughly, HMM-RPF performs worst.

A running time comparison among the involved algorithms is presented in Table \ref{Table:time}. It shows that, for case $P_o=0.1$, DPM-RPF has a computational complexity in between ILAPF and HMM-RPF; while for case $P_o=0.9$, the running time of DPM-RPF becomes larger than the others. We can obtain the reason of this result by performing an analysis on the complexity cost of DPM-RPF. Due to the presence of the DPM outlier modeling procedure, as more outliers appear, the complexity of the algorithm will be increased accordingly.
\begin{table}\centering\small
\caption{The mean running time (in seconds) calculated over 100 independent runs for cases $P_o=0.1$ and $P_o=0.9$}
\begin{tabular}{c|c|c|c}
\hline %
Algorithm & HMM-RPF & ILAPF & DPM-RPF \\\hline
Case $P_o=0.1$ & 14.0053 & 7.3357 & 10.8466 \\\hline
Case $P_o=0.9$ & 13.2726 & 6.9697 & 30.9837 \\\hline
\end{tabular}
\label{Table:time}
\end{table}
\section{Conclusions}\label{sec:conclusion}
In this paper, we presented a Bayesian nonparametrics based robust PF algorithm termed DPM-RPF. We applied the DPM model to characterize the unknown generative mechanism of the outliers and then derived the DPM-RPF algorithm for sequential posterior inference of the outlier model as well as the system state of interest.

The experimental result provides a strong evidence on the superiority of the presented algorithm in terms of discovering the mixture patterns underlying the outliers. It also shows that the more frequently the outliers appear, the more obvious the advantage of DPM-RPF in terms of filtering accuracy. The complexity cost of the proposed algorithm is empirically studied (see Table I). It is shown that the complexity cost of DPM-RPF is dependant on the number of outliers. As outliers appear more frequently, the computation complexity of DPM-RPF increases accordingly, and vice versa.

A further rigorous theoretical study and more realistic application studies in scenarios like multi-target tracking in clutter \cite{liu2010multi} and filtering with imprecisely time-stamped measurements \cite{millefiori2015adaptive} can be conducted as future work. In addition, how to configure hyper-parameters of the model in a smarter way is also interesting to be investigated.
\bibliographystyle{IEEEbib}
\bibliography{mybibfile}
\end{document}